\documentclass[10pt,twocolumn,letterpaper]{article}

\usepackage{cvpr}
\usepackage{times}
\usepackage{epsfig}
\usepackage{graphicx}
\usepackage{amsmath}
\usepackage{amssymb}
\usepackage{float}
\usepackage{subfig}
\usepackage{booktabs}
\usepackage{multirow}
\usepackage{array}
\usepackage{color}
\usepackage{tabularx}
\usepackage{longtable}
\usepackage{tabu}
\usepackage{algorithmic}
\usepackage[ruled,linesnumbered]{algorithm2e}

\usepackage{booktabs}

\DeclareMathOperator*{\argmax}{argmax}

\usepackage[breaklinks=true,bookmarks=false]{hyperref}

 \cvprfinalcopy % *** Uncomment this line for the final submission

\def\cvprPaperID{11775} % *** Enter the CVPR Paper ID here

\begin{document}

%%%%%%%%% TITLE
\title{Debiased Subjective Assessment of Real-World Image Enhancement}

\author{Peibei Cao$^1$, Zhangyang Wang$^2$, and Kede Ma$^1$\\
$^1$ City University of Hong Kong, $^2$ University of Texas at Austin\\
{\tt\small peibeicao2-c@my.cityu.edu.hk, atlaswang@utexas.edu, 
kede.ma@cityu.edu.hk}
% For a paper whose authors are all at the same institution,
% omit the following lines up until the closing ``}''.
% Additional authors and addresses can be added with ``\and'',
% just like the second author.
% To save space, use either the email address or home page, not both
%\and
% Second Author\\
% Institution2\\
% First line of institution2 address\\
% {\tt\small secondauthor@i2.org}
}

\pagestyle{empty}  % no page number for the second and the later pages
\thispagestyle{empty}
\maketitle
\thispagestyle{empty}

%%%%%%%%% ABSTRACT
\begin{abstract}
% For many years, image enhancement  has been investigated in an unrealistic setting, where the performance is quantified using the average distance between a set of enhanced  and corresponding ``ideal'' images. However,
In real-world image enhancement, it is often challenging (if not impossible) to acquire ground-truth data, preventing the adoption of distance metrics for objective quality assessment. As a result, one often resorts to subjective quality assessment, the most straightforward and reliable means of evaluating image enhancement. Conventional subjective testing requires manually pre-selecting a small set of visual examples, which may suffer from three sources of biases: 1) sampling bias due to the extremely sparse distribution of the selected samples in the image space; 2) algorithmic bias due to potential overfitting the selected samples; 3) subjective bias due to  further potential cherry-picking test results. This eventually makes the field of real-world image enhancement more of an art than a science. Here we take steps towards debiasing conventional subjective assessment by automatically sampling a set of adaptive and diverse images for subsequent testing. This is achieved by casting sample selection into a joint maximization  of the discrepancy between the enhancers and the diversity among the selected input images.
Careful visual inspection  on the resulting enhanced images provides a debiased ranking of  the enhancement algorithms. We demonstrate our subjective assessment method using  three popular and practically demanding image enhancement tasks: dehazing, super-resolution, and low-light enhancement.
\end{abstract}

%%%%%%%%% BODY TEXT
\section{Introduction}
For many years, image enhancement has been investigated in an unrealistic setting, with the assumption that the original images of perfect quality exist to help evaluate visual quality of the enhanced images. This promotes the use of full-reference image quality metrics \cite{wang2006modern} to compute an average distance between a large set of enhanced and original image pairs as an indication of enhancement performance. Along this path, many full-reference metrics have been proposed \cite{wang2004image,sheikh2006image,zhang2018unreasonable,ding2020image}, trying to measure this distance more perceptually. 

However, in real-world image enhancement, it is often difficult (if not impossible) to specify desired outputs. Moreover, there may be multiple diverse outputs that are desirable, as in the case of super-resolution \cite{yang2010image}. Therefore, full-reference models that rely on a single  ``ideal'' image are not applicable. Some attempts have been made to adopt no-reference 
models \cite{wang2011reduced} for performance assessment of real-world enhancement. However, no-reference objective assessment is still in its infancy, and accurate models for (specific or general) image enhancement applications are largely lacking. Currently, the most widely used no-reference metric - NIQE \cite{mittal2012making} - was empirically proven to correlate poorly with human quality judgments of the enhanced images \cite{mittal2012making}, which exhibit unique and algorithm-specific artifacts that are often non-overlapping with natural distortions.

Alternatively, one may refer to subjective quality assessment, which is so far the most straightforward and  reliable way of evaluating real-world image enhancement because the ultimate receiver in most such applications is the human eye. Conventional subjective assessment typically takes a \textit{four-step approach}. First, pre-select a number of images from the input domain of a given  image enhancement problem. Second, pick a set of competing enhancers, and generate the corresponding output images. Third, ask humans to rate the perceived quality of the enhanced images. Fourth, compare the enhancers according to the subjective results. 

Unfortunately, conventional subjective assessment may suffer from three sources of biases. The first is the \textit{sampling bias}.
The underlying principle of conventional subjective assessment is to prove an enhancement method to be correct. This would require the set of pre-selected images to be large enough to sufficiently represent the input domain of interest. However, subject testing is an expensive and time-consuming endeavor. In practice, the number of images being examined is limited to a few hundreds (if not fewer), casting doubt on the assumption of sufficient sampling in the high-dimensional image space. The second is the \textit{algorithmic bias}.  It is important to note that the selection of test images precedes the selection of competing methods. One may take advantage of this (intentionally or unintentionally), and tunes her/his enhancer to overfit the pre-selected images, drawing overly optimistic conclusions on the real-world generalization performance. The third is the \textit{subjective bias}. That is, the test results may further be cherry-picked to bias towards certain methods. In summary, it is sad, but not uncommon, to see that a ``state-of-the art'' image enhancer produces superior results in its original paper, but remains particularly weak at handling examples appeared in subsequent work. 

In this paper, we contribute to debiasing conventional  subjective assessment by injecting an \textit{automated},  \textit{adaptive} and \textit{sample-efficient} mechanism to select input domain samples. Our inspirations are drawn from interdisciplinary prior work on ``model falsification as model comparison", a renowned philosophy in the fields of computational vision \cite{wang2008maximum}, software testing \cite{mckeeman1998differential} and computer vision \cite{pei2017deepxplore,ma2018group,wang2020going}. Specifically, we start from a large-scale image set as a finite approximation to the input space of an image enhancement application. According to the available human labelling budget, our method automatically selects a set of adaptive and diverse images for subsequent subjective testing. The selected images are optimal in terms of discriminating between the enhancers, while having the maximum within-group variation in a latent space to ensure content diversity. Subjective results of the corresponding enhanced images reveal the advantages and disadvantages of the competing methods, and provide a debiased ranking of their relative performance. Our subjective assessment method is applicable to a wide variety of image processing and computer photography subfields, and we choose three real-world image enhancement applications as demonstration: 1) single image dehazing, 2) single image super-resolution, and 3) low-light image enhancement.

\section{Related Work}
We provide a concise overview of three real-world image enhancement applications,   
%single image dehazing, single image super-resolution, and low-light image enhancement, 
with emphasis on how previous subjective and objective assessments were carried out. 

\vspace{-0.5em}
\paragraph{Single Image Dehazing} Outdoor images are often captured in the presence of haze \cite{narasimhan2002vision}. Due to the absorption and refraction of light by turbid medium, the resulting images may suffer from poor visibility and color shift.  Conventional single image dehazing methods relied on the Koschmieder's model \cite{harald1924theorie} and natural image priors \cite{Hautiere2007dehaze,He2010dehaze,Meng2013dehaze,zhu2015dehaze,Berman2016dehaze} to estimate the atmospheric light and the transmission map.  With the recent advances in convolutional neural networks (CNNs), plenty of CNN-based methods ~\cite{zhang2018dehaze,Ren2018dehaze,li2017aod,Ren2016dehaze,Cai2016dehaze} have been proposed, directly regressing clean images from hazy ones. 

Ma \etal~\cite{MaDehaze} made initial attempts to subjective assessment of single image dehazing. A somewhat surprising observation is that due to the introduction of algorithm-dependent distortions, the dehazed results by some algorithms are statistically insignificant compared to the input hazy images. Choi \etal~\cite{ChoiDehaze} put emphasis on perceived fog density instead of overall quality. Li \etal~\cite{Li2019} evaluated several dehazing algorithms for both human and machine vision. 
Tang \etal~\cite{TangDehaze} investigated nighttime image dehazing, asking subjects to rate  four aspects of dehazed algorithms: detail recognition, color fidelity, image authenticity, and overall effect. 

The above subjective studies lead to an increasing consensus that objective quality models such as the mean squared error (MSE), the structural similarity (SSIM) index \cite{wang2004image} and other no-reference methods \cite{hautiere2008blind} cannot accurately predict the perceived quality of dehazed images.

\vspace{-0.5em}
\paragraph{Single Image Super-Resolution} Super-resolving a low-resolution image into a high-resolution one is very challenging, especially with a large scaling factor. Early attempts were mainly interpolation-based methods~\cite{Zhang2006SR} using natural image statistics. In the late 2000s,  model-based methods~\cite{Dong2011SR} came into play, with gradient profile prior \cite{sun2008image}, sparsity prior~\cite{yang2010image}, and self-similarity prior~\cite{huang2015single} being representative. In 
the past five years, CNN-based methods began to dominate this field~\cite{7115171,Kim2016enlighten,Lai2017enlighten,zhang2018learning,zhang2018rcan,zhang2018Residual}, some of which were combined with generative adversarial networks (GANs) to encourage texture synthesis \cite{zhu2017unpaired}. 

Yang \etal~\cite{YangSR} presented one of the first subjective evaluations of single image super-resolution methods. Later, the authors~\cite{MaSR} enlarged their dataset by including more high-resolution images and more competing methods.  Johnson \etal~\cite{JohnsonSR} performed a small-scale subjective experiment to verify the perceptual advantages of the VGG loss \cite{Simonyan14c} in super-resolution. Gu \etal \cite{gu2020pipal} established PIPAL - a large-scale subject-rated dataset for image restoration, including GAN-based super-resolution results. With respect to objective assessment, MSE and SSIM~\cite{wang2004image} are still the most widely adopted metrics. LPIPS \cite{zhang2018unreasonable} and DISTS \cite{ding2020image} that give credit to visually plausible synthesized textures have also been used for benchmarking purposes in some of the latest work. When high-resolution images are assumed unknown, NIQE is sometimes used for quantitative comparison.

\begin{figure*}[t]
\centering
\vspace{-.5cm}
\includegraphics[height=0.5\linewidth]{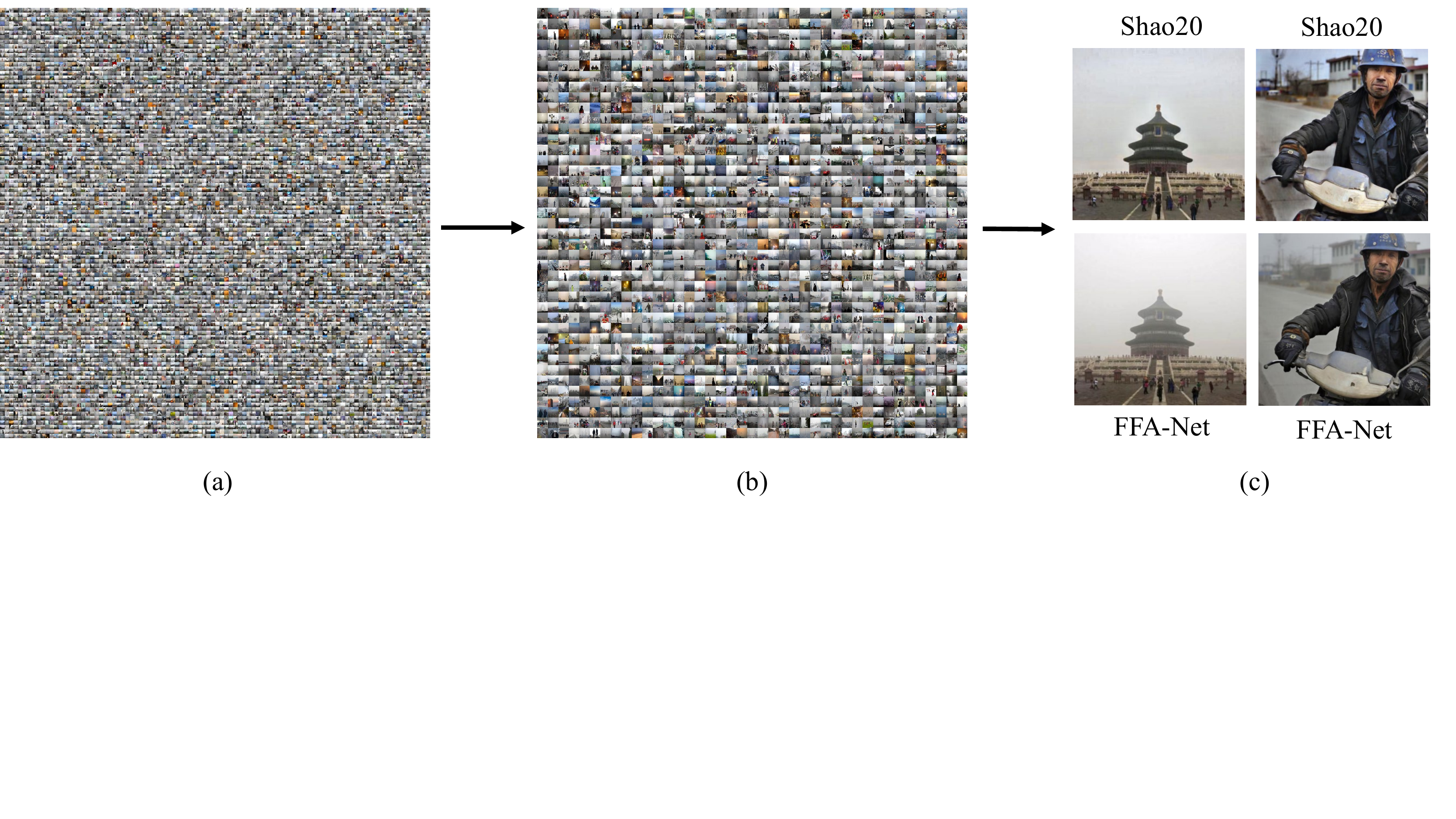}
\vspace{-3.5cm}
   \caption{Overview of our debiased subjective assessment in the context of single image dehazing. 
       \textbf{(a)}: A large set of hazy images as an approximation to the input domain $\mathcal{X}$. \textbf{(b)}: Top-$K$ hazy images selected from (a) to best discriminate between Shao20 \cite{Shao2020} and FFA-Net \cite{FFA-net} by optimizing Eq. \eqref{eq:mad2}. $D_1$ and $D_2$ are implemented by DISTS \cite{ding2020image} and MSE of the last feature layer of VGGNet \cite{Simonyan14c}, respectively.  \textbf{(c)}: Pairs of dehazed images corresponding to representative hazy images in (b).}
   \vspace{-0.5em}
\label{fig:dataset}
\end{figure*}

\vspace{-0.5em}
\paragraph{Low-Light Image Enhancement}Arguably the most significant impediment to high-quality pictures is lack of light \cite{hasinoff2016burst}. On the one hand, a nighttime or indoor scene may not provide adequate light. On the other hand, a daytime scene may has a high dynamic range (\ie, unbalanced lighting), causing current imaging techniques to collect insufficient light in shadow regions. Early computational methods for low-light image enhancement were equated to contrast enhancement either globally~\cite{Coltuc2006enlighten} or locally~\cite{Stark2000enlighten}. The Retinex theory~\cite{Edwin1977enlighten} was also extensively studied in this context, where the problem of low-light enhancement is transformed to illumination map estimation \cite{guo2016lime}. Recently, many data-driven CNN-based methods~\cite{Jiang2019enlighten,Chen2018Retinex,Rui2019enlighten,Jiang2019enlighten} with and without paired supervision have been developed, obtaining superior results on a limited number of visual examples. 

 Limited work has been done to assess low-light image enhancement subjectively.  Hwang \etal~\cite{HwangEnhance} carried out a user study to validate their proposed enhancer using $20$ low-contrast images, some of which are due to poor lighting conditions. Chen \etal~\cite{chen2014quality} included images captured in hazy, underwater, and low-light conditions for human testing. 
 A recent subjective study~\cite{Zero-DCE} compared six advanced low-light enhancers. Another small-scale subjective study was reported in \cite{Jiang2019enlighten} on 23 low-light images with six enhancement algorithms. With regard to objective assessment, MSE and SSIM prevail in this application. Using the input image as a corrupted reference, one may refer to VIF \cite{sheikh2006image} and PCQI \cite{wang2015patch} for measuring the degree of enhancement. To the best of our knowledge, existing no-reference models \cite{fang2014no} remain particularly weak at predicting the perceived quality of low-light enhanced images.

 The above-mentioned subjective experiments may differ in how test images are presented to the subjects and how human data are collected, but they all need to pre-select the test images by the experimenters. Therefore, the results may suffer from sampling, algorithmic, and subjective biases, motivating us to debias subjective assessment of real-world image enhancement in this work.

\section{Proposed Method}
% The general problem of model comparison for evaluating CP techniques may be formulated as follows.
 We formulate subjective assessment of real-world image enhancement in a general mathematical framework. Starting from an input image domain $\mathcal{X}$, we can easily sample an image $x\in \mathcal{X}$ that needs to be enhanced for improved visual quality. We choose a set of enhancement methods $\mathcal{F}=\{f_j\}_{j=1}^N$, each of which takes an $x\in\mathcal{X}$ as input, and produces an enhanced output $y_j = f_j(x)$. We also assume a subjective assessment environment, where human participants can reliably rate the perceived quality of $y_j$. The ultimate goal is to compare the $N$ methods on the input domain $\mathcal{X}$ containing enormous images, under the constraint of a very limited human labelling budget.
 
Conventional subjective assessment first pre-selects a small image set $\mathcal{S}=\{x^{(i)}\}_{i=1}^M$. For each image $x\in\mathcal{S}$, a set of enhanced versions $\{y_j\}_{j=1}^N$ are created, based on which subjective testing reveals the relative performance of $\{f_j\}_{j=1}^N$ on $x$. 
% Active sampling may be adopted at this stage to reduce the labelling cost \cite{Ye2014,Fan2017,Li2018}. 
The model with the highest subjective ratings averaged over $\mathcal{S}$ is the best. As discussed previously, this would  introduce several sources of biases.
Inspired by interdisciplinary work under the scientific philosophy of ``model falsification as model comparison'' \cite{mckeeman1998differential,pei2017deepxplore,ma2018group}, especially following the well-established principle of maximum differentiation (MAD) competition~\cite{wang2008maximum}, we aim to falsify an enhancer by finding a minimum set of images, which are most likely to be its counterexamples. \textit{An enhancer that is more difficult to be falsified is considered better}.

We first describe the simplest situation, where two enhancers $f_1$ and $f_2$ are being compared, and the human labelling budget only allows us to select a single image $x\in \mathcal{X}$ for subjective testing. Then, the core question boils down to: 
How to automatically select which image for subjective testing from massive candidate images, such that the relative performance $f_1$ and $f_2$ may be most easily revealed? 

\begin{figure}[t]
\centering
\addtocounter{subfigure}{-1}
\subfloat{\includegraphics[width=0.32\linewidth]{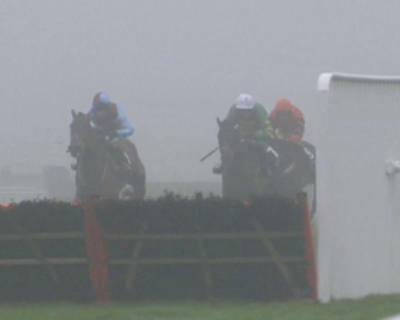}}\hskip.3em
\subfloat[Without $D_2$]{\includegraphics[width=0.32\linewidth]{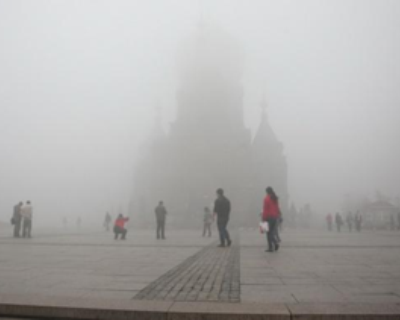}}\hskip.3em
\subfloat{\includegraphics[width=0.32\linewidth]{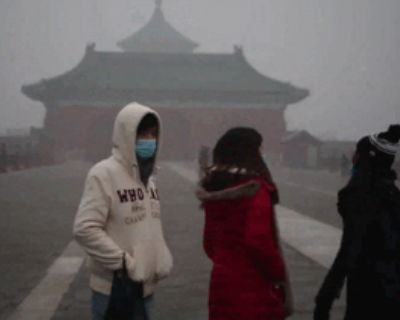}}
\\
\vspace{-0.2cm}
\addtocounter{subfigure}{-2}
\subfloat{\includegraphics[width=0.32\linewidth]{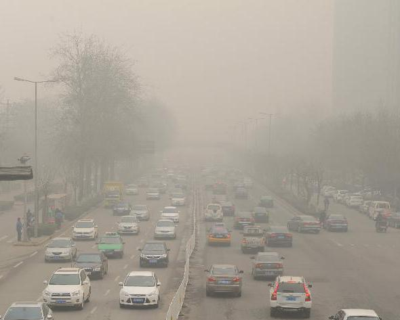}}\hskip.3em
\subfloat[With $D_2$]{\includegraphics[width=0.32\linewidth]{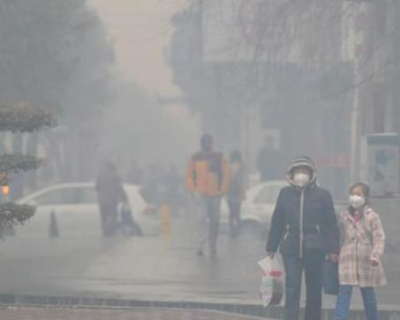}}\hskip.3em
\subfloat{\includegraphics[width=0.32\linewidth]{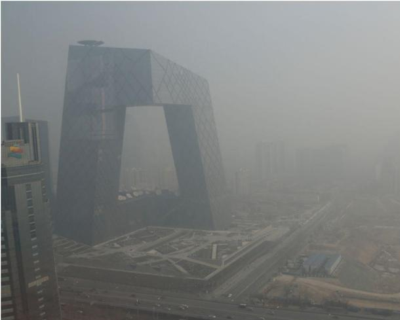}}

% \fbox{\rule{0pt}{2in} \rule{.9\linewidth}{0pt}}
\vspace{-.3cm}
   \caption{Top-$K$ images selected (a) without and (b) with the diversity loss, respectively.}
   \vspace{-0.3cm}
\label{fig:diversity}
\end{figure}

According to the MAD competition methodology~\cite{wang2008maximum}, our method selects the image $\hat{x}\in \mathcal{X}$ that best differentiates between $f_1$ and $f_2$:
 \begin{align}\label{eq:mad}
    \hat{x} = \argmax_{x \in \mathcal{X}}  D_1(f_1(x), f_2(x)),
\end{align}
where $D_1$ is a quantitative measure to approximate the perceptual distance between $f_1(x)$ and $f_2(x)$. Visual inspection on $f_1(\hat{x})$ and $f_2(\hat{x})$ leads to two plausible results:
\begin{itemize}
    \item The majority of human subjects prefer $f_1(\hat{x})$ (or $f_2(\hat{x})$) over $f_2(\hat{x})$ (or $f_1(\hat{x})$). In this case, the proposed subjective assessment method automatically detects a strong counterexample of one enhancer, not the other; a clear winner is declared. The chosen $\hat{x}$ is the most informative in ranking the relative performance between $f_1$ and $f_2$.
    \item Human subjects give $f_1(\hat{x})$ and $f_2(\hat{x})$  similar ratings. High rating indicates that both methods generate desirable but diverse outputs, which makes sense  in real-world image enhancement that admits multiple plausible outputs. Low rating indicates that both fail, in dramatically different ways, to produce reasonable results. In either case, the chosen $\hat{x}$ 
   reveals different aspects of the strengths (or weaknesses) of $f_1$ and $f_2$, but contributes less to their relative performance ranking.
\end{itemize}
It seems straightforward to extend this idea to compare $f_1$ and $f_2$ on a small image subset $\mathcal{S}\subset\mathcal{X}$ containing images with top-$K$ largest distances computed by Eq. \eqref{eq:mad}. However, such a na\"{i}ve extension may simply identify algorithm failures of the same underlying root cause, leading to less interesting comparison (see Figure \ref{fig:diversity}).
To encourage more diverse failures of the competing models to be spotted, 
we modify Eq. \eqref{eq:mad} when looking for the $k$-th image:
 \begin{align}\label{eq:mad2}
    \hat{x}^{(k)} = \argmax_{x \in \mathcal{X}\setminus\mathcal{S}} ~ D_1(f_1(x), f_2(x)) +\lambda_1 D_2(x,\mathcal{S}),
\end{align}
where $\mathcal{S}=\{\hat{x}^{(i)}\}_{i=1}^{k-1}$ is the set of $k-1$ images that have already been identified according to Eq. \eqref{eq:mad2}. $D_2$ is a second measure to quantify the semantic distance between an image $x$ and the set $\mathcal{S}$. $\lambda_1$ governs the trade-off between the two terms. Once $\hat{x}^{(k)}$ is identified, we incorporate it into $\mathcal{S}$.

\begin{algorithm}[t]
\caption{Debiased Subjective Assessment of Real-World Image Enhancement}
\label{alg:Framwork} 
\KwIn{A large-scale set $\mathcal{X}$, a list of competing methods $\mathcal{F}=\left\{f_{j}\right\}_{j=1}^{N}$, and two distance measures $D_1$ and $D_2$\\}
\KwOut{Global ranking vector $\mu \in \mathbb{R}^{N}$}
$\mathcal{D}\leftarrow\emptyset$

\For{$j \gets 1$ \KwTo $N$}
{
Compute the enhanced images $\left\{f_{j}(x)\vert x \in \mathcal{X}\right\}$
}
\For{$i \gets 1$ \KwTo $N-1$}
{
\For{$j \gets i+1$ \KwTo $N$}
{
$\mathcal{S}\leftarrow\emptyset$

\For{$k \gets 1$ \KwTo $K$}
{
Select $\hat{x}^{(k)}$ by optimizing Eq. \eqref{eq:mad2}

$\mathcal{S}\leftarrow\mathcal{S}\cup\hat{x}^{(k)}$

$\mathcal{D}\leftarrow\mathcal{D}\cup\{f_i(\hat{x}^{(k)}), f_j(\hat{x}^{(k)})\}$
}
\label{alg:Framwork:result}
% Switch the roles of $f_{i}$ and $f_{j}$, where $f_{i}$ and $f_{j}$ are the attacker and the defender, respectively, and repeat Step \ref{alg:Framwork:com} to Step \ref{alg:Framwork:result}
}
}

Create the count matrix $C$ for $\mathcal{D}$ via the 2AFC method

Compute  $\mu$ by optimizing Eq. \eqref{eq:likelihood}
\end{algorithm}

\begin{figure*}[t]
\centering
% \includegraphics[height=0.55\linewidth]{fig/ranking/rank.pdf}
% \subfloat[]{\includegraphics[height=0.49\linewidth]{fig/ranking/dehaze1.pdf}}\hspace{-6mm}
% \subfloat[]{\includegraphics[height=0.49\linewidth]{fig/ranking/enhance1.pdf}}\hspace{-6mm}
% \subfloat[]{\includegraphics[height=0.49\linewidth]{fig/ranking/SR1.pdf}}\hskip.3em
\subfloat[]{\includegraphics[height=0.3\linewidth]{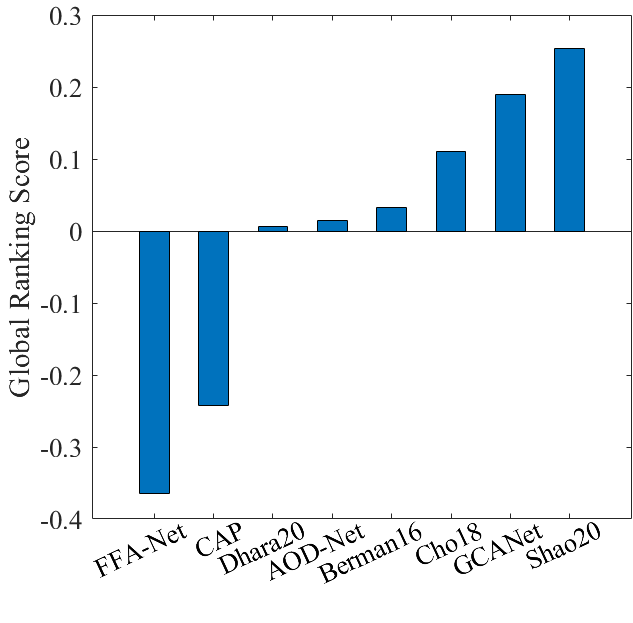}}\hskip.1em
\subfloat[]{\includegraphics[height=0.3\linewidth]{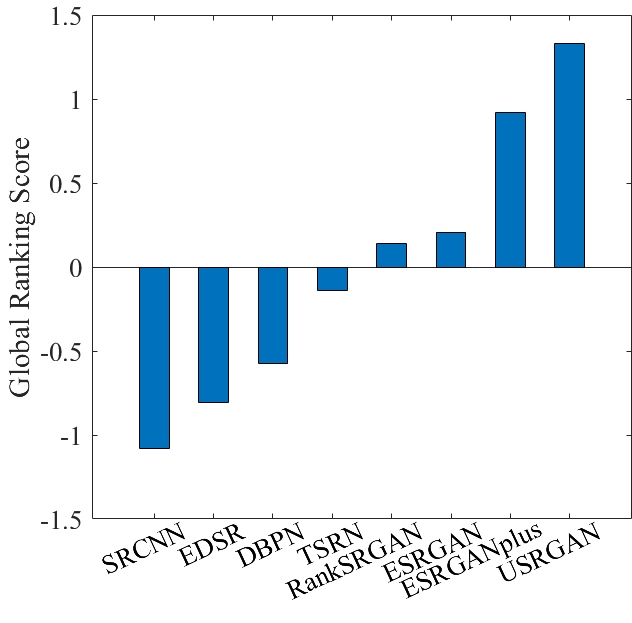}}\hskip.1em
\subfloat[]{\includegraphics[height=0.3\linewidth]{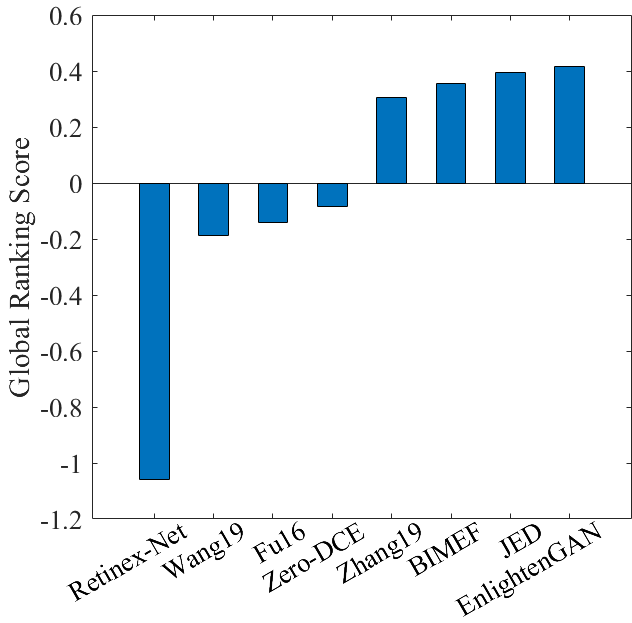}}\hskip.1em
\vspace{-0.3cm}
   \caption{Global ranking results of  
       \textbf{(a)} single image dehazing, \textbf{(b)} single image super-resolution, and \textbf{(c)} low-light image enhancement by optimizing Eq. \eqref{eq:likelihood}.}
   \vspace{-0.3cm}
\label{fig:ranking}
\end{figure*}

Given $N$ enhancement algorithms, our subjective assessment method chooses top-$K$ images for each of the $\binom{N}{2}$ distinct pairs of enhancers, gives rise to a final set $\mathcal{D}$ with  $N(N-1)K$ enhanced images. It is worth noting that the size of $\mathcal{D}$ is independent of the size of the input domain $\mathcal{X}$. Provided that the computational cost of image enhancement is negligible, the proposed subjective assessment method suggests expanding $\mathcal{X}$ to cover as many images (and therefore diverse failures of the competing methods) as possible.

We now introduce the subjective assessment environment for gathering human opinions of image quality. As  each input image $x\in \mathcal{S}$ is associated with a pair of enhanced images $\{f_i(x), f_j(x)\}\subset \mathcal{D}$, it is natural to employ the two-alternative forced choice (2AFC) method. That is, the subject is presented with $f_i(x)$ and $f_j(x)$ simultaneously, and is forced to choose which one is of higher quality.  After subjective testing, we arrange the collected human data in an $N\times N$ matrix $C$, where $C_{ij}$ records the number of votes for $f_i$ and against $f_j$. Finally, we adopt maximum likelihood for multiple options \cite{tsukida2011analyze} under the Thurstone's model \cite{thurstone1927law} to infer the global ranking of $\mathcal{F}$. Specifically, we let $\mu$ be the vector of global ranking scores $[\mu_1,\mu_2,\ldots,\mu_N]$, and define the log-likelihood of the count matrix, $C$, as 
\begin{align}\label{eq:likelihood}
    L(\mu;C)=\sum_{ij}C_{ij}\log( \Phi(\mu_i-\mu_j)),
\end{align}
where $\Phi(\cdot)$ is the standard Normal cumulative distribution function. When maximizing $L(\mu;C)$, one often adds an addition constraint, $\sum_i \mu_i=0$, to obtain a unique optimal solution. We summarize the proposed debiased subjective assessment method in Algorithm \ref{alg:Framwork}, and show an overview of it in the context of single image dehazing in Figure \ref{fig:dataset}.

% , and show the overview of the debiased subjective assessment in the context of single image dehazing in Figure \ref{fig:dataset}

% \begin{figure}[t]
% \begin{center}
% \fbox{\rule{0pt}{2in} \rule{0.9\linewidth}{0pt}}
%   %\includegraphics[width=0.8\linewidth]{egfigure.eps}
% \end{center}
%   \caption{Example of caption.  It is set in Roman so that mathematics
%   (always set in Roman: $B \sin A = A \sin B$) may be included without an
%   ugly clash.}
% \label{fig:long}
% \label{fig:onecol}
% \end{figure}

% \begin{figure*}
% \begin{center}
% \subfloat[]{\includegraphics[height=0.2\linewidth]{fig/ranking/dehazing.png}}\hskip.3em
% \vspace{-0.2cm}
% \subfloat[]{\includegraphics[height=0.2\linewidth]{fig/ranking/enhance.png}}\hskip.3em
% \vspace{-0.2cm}
% \subfloat[]{\includegraphics[height=0.2\linewidth]{fig/ranking/SR.png}}
% % \fbox{\rule{0pt}{2in} \rule{.9\linewidth}{0pt}}
% \end{center}
% \vspace{-.2cm}
%   \caption{(a)-(c) are the group ranking and t-test results for dehazing, low-light enhancement and SR respectively.}
% %   \vspace{-.3cm}
% \label{fig:ttest1}
% \end{figure*}

\section{Applications to Image Enhancement}
In this section, we apply our subjective assessment method to three real-world image enhancement tasks: single image dehazing, single image super-resolution, and low-light image enhancement.

\subsection{Experimental Setups}\label{subsec:setup}

\paragraph{Construction of $\mathcal{X}$} For dehazing, the $10,000$ real hazy images are originated from RESIDE \cite{liu2018dehaze} and the Internet. For super-resolution, the $10,000$ low-resolution images are from WED~\cite{ma2016waterloo}, OST~\cite{wang2018sftgan}, and the Internet. For low-light enhancement, the $10,000$ low-light images are chosen from ExDark~\cite{loh2019getting}, NPE~\cite{wang2013naturalness}, DICM~\cite{lee2012contrast}, MBLLEN~\cite{lv2018mbllen}, VV~\cite{vonikakis2018evaluation}, and the Internet. No manual pre-screening is needed at this stage.

\vspace{-0.5em}
\paragraph{Selection of $\mathcal{F}$} For dehazing, we select eight popular algorithms published from 2015 to 2020: CAP~\cite{zhu2015dehaze}, Berman16~\cite{Berman2016dehaze}, AOD-Net~\cite{li2017aod}, Cho18~\cite{Cho18}, GCANet \cite{GCA}, FFA-Net \cite{FFA-net}, Dhara20 \cite{Dhara2020}, and Shao20 \cite{Shao2020}, among which CAP, Berman16, Cho18, and Dhara20 are knowledge-driven, while the rest are data-driven.

For super-resolution, we select eight CNN-based methods ranging from 2016 to 2020: SRCNN~\cite{7115171}, EDSR~\cite{EDSR}, DBPN~\cite{DBPN}, TSRN~\cite{Texturenet}, ESRGAN~\cite{ESRGAN}, RankSRGAN~\cite{RankSRGAN}, ESRGANplus~\cite{ESRGAN+}, and 
USRGAN~\cite{zhang2020deep}. 

For low-light enhancement, we select  eight methods from 2016 to 2020: Fu16~\cite{Fu2016fussion}, BIMEF~\cite{ying2017bio}, Retinex-Net~\cite{Chen2018Retinex}, JED~\cite{Ren2018JED}, EnlightenGAN~\cite{Jiang2019enlighten}, Zhang19~\cite{Zhang2019dual}, Wang19~\cite{Wang2019ad}, and Zero-DCE~\cite{Zero-DCE}, among which Retinex-Net, EnlightenGAN, and Zero-DCE are CNN-based. The implementations of all 24 methods are obtained from the respective authors, and are tested with the default settings.

\vspace{-0.5em}
\paragraph{Construction of $\mathcal{S}$} The created $\mathcal{X}$ may be noisy, including images that lie out of the input domain of interest. Therefore, for dehazing, we replace a selected image that is either non-natural or haze-free with the next eligible one that optimizes Eq. \eqref{eq:mad2}. Moreover, for each $x\in \mathcal{S}$, the visibility improvements in the corresponding ``dehazed'' images $f_i(x)$ and $f_j(x)$ are automatically checked by the computational method in \cite{ChoiDehaze}. If there is no predicted improvement in either dehazed image, we discard $x$. We apply the same image screening procedure for low-light enhancement, where the computational method in \cite{fang2014no} is adopted for automatic verification of detail enhancement.  

% FIRST-ORDER HEADINGS. (For example, {\large \bf 1. Introduction})
% SECOND-ORDER HEADINGS. (For example, { \bf 1.1. Database elements})

\vspace{-0.5em}
%-------------------------------------------------------------------------
\paragraph{Subjective Experiment}
We conduct subjective user studies to gather human quality scores of the enhancement results in $\mathcal{D}$. The 2AFC method is adopted, allowing  differentiation of subtler quality variations. Subjects are forced to choose the image with higher perceived quality with unlimited viewing time. For each enhancement application, we set $K=12$, resulting in a total of $\binom{8}{2}\times 12=336$ paired comparisons. To relieve fatigue,  subjects are allowed to take a break at anytime during each session of subjective testing.  We gather data from $25$ subjects with general background knowledge of image processing.
%-------------------------------------------------------------------------
%-------------------------------------------------------------------------

% {\small\begin{verbatim}
%   \usepackage[dvips]{graphicx} ...
%   \includegraphics[width=0.8\linewidth]
%                   {myfile.eps}
% \end{verbatim}
% }
\begin{figure*}
\begin{center}
\subfloat[Dhara20/Shao20]{\includegraphics[height=0.165\linewidth]{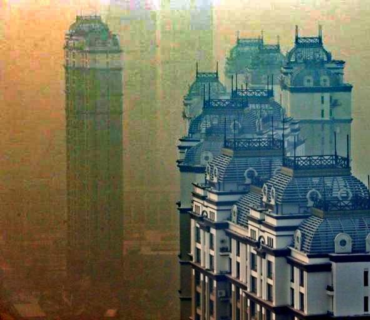}}\hskip.3em
\subfloat[Dhara20/Cho18]{\includegraphics[height=0.165\linewidth]{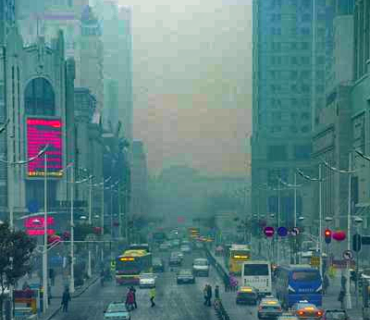}}\hskip.3em
\subfloat[Berman16/Cho18]{\includegraphics[height=0.165\linewidth]{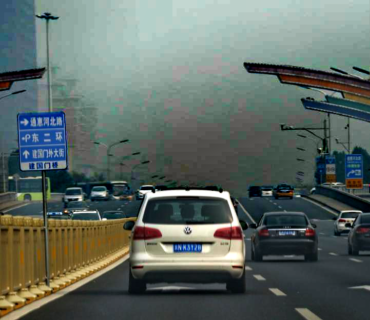}}\hskip.3em
\subfloat[Cho18/Shao20]{\includegraphics[height=0.165\linewidth]{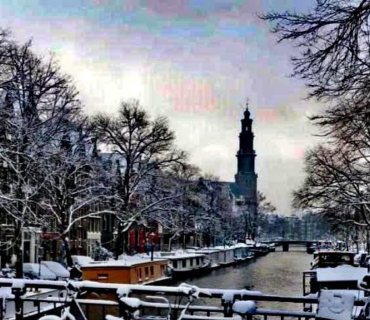}}\hskip.3em
\subfloat[GCANet/AOD-Net]{\includegraphics[height=0.165\linewidth]{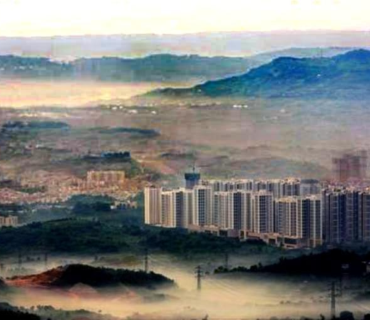}}\hskip.3em
\\
\vspace{-0.2cm}
\subfloat[GCANet/CAP]{\includegraphics[height=0.165\linewidth]{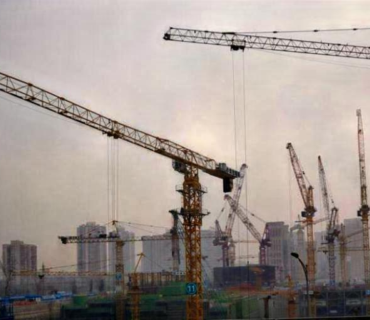}}\hskip.3em
\subfloat[Shao20/Cho18]{\includegraphics[height=0.165\linewidth]{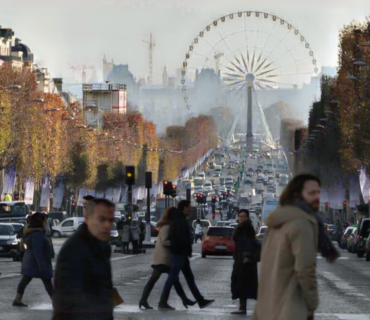}}\hskip.3em
\subfloat[CAP/GCANet]{\includegraphics[height=0.165\linewidth]{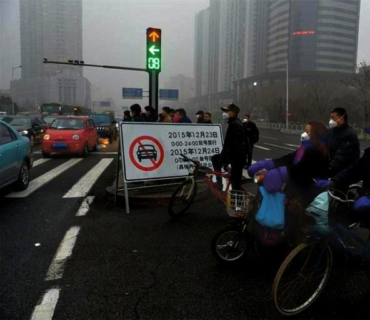}}\hskip.3em
\subfloat[AOD-Net/Shao20]{\includegraphics[height=0.165\linewidth]{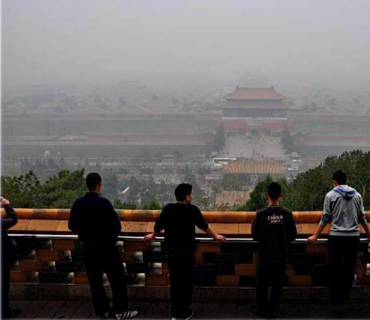}}\hskip.3em
\subfloat[FFA-Net/Berman16]{\includegraphics[height=0.165\linewidth]{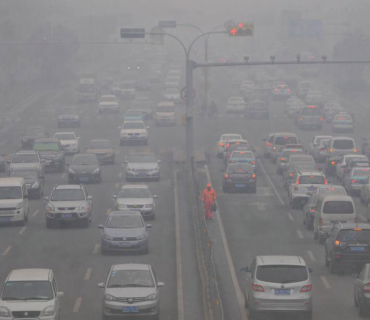}}\hskip.3em

% \fbox{\rule{0pt}{2in} \rule{.9\linewidth}{0pt}}
\end{center}
\vspace{-.5cm}
  \caption{Representative distortions created by dehazing methods in our experiment. $f_i/f_j$ below each image means that $f_i$ is used to produce the image, while $f_j$ is the paired method for selecting the corresponding hazy image in Eq. \eqref{eq:mad2}.}
  \vspace{-0.4cm}
\label{fig:dehazing}
\end{figure*}

%-------------------------------------------------------------------------

\subsection{Main Results}
\paragraph{Quantitative Results}
 We show the global ranking results of the three real-world image enhancement applications in Figure \ref{fig:ranking}, from which we have several interesting observations.
 
 \vspace{0.5em}
 For \textit{dehazing}, the main observation is that the synthetic-to-real domain shift challenges all methods. Shao20 \cite{Shao2020} leverages a bidirectional network to explicitly bridge the gap between the synthetic and realistic hazy images, and therefore exhibits the strongest generalization to the real world. By contrast, FFA-Net \cite{FFA-net} relies exclusively on synthetic data for training, the majority of which are indoor scenes. Along with delicate feature attention and fusion modules, FFA-Net tends to overfit synthetic data, and has the worst performance in the debiased subjective experiment. Second, methods with less reliance on the Koschmieder’s model~\cite{harald1924theorie} and image priors, such as GCANet \cite{GCA} and Cho18 \cite{Cho18},  generally outperform prior-based methods Berman16 \cite{Berman2016dehaze}, Dhara20 \cite{Dhara2020}, and CAP~\cite{zhu2015dehaze}, and the physical model-based AOD-Net \cite{li2017aod}. This makes sense because current physical and statistical models oversimplify the natural imaging process in complex realistic hazy scenes, \eg, in the presence of non-uniform/dark light or heterogeneous haze density. As a result, algorithm-dependent artifacts are likely to emerge, which may be perceptually more annoying than the haze (see Figure \ref{fig:dehazing}). Last, the Spearman's rank correlation coefficient (SRCC) between the subjective results of the competing methods and their publication times is only $0.167$, implying the progress made in the field of single image dehazing might be somewhat over-estimated in terms of their real-world generalization, despite outstanding (synthetic) benchmark numbers.
 
% The overstated performance in the field of single image dehazing has been pointed out by our subjective assessment method.

For \textit{super-resolution}, steady progress over the years has been reported in our experiment, with an SRCC value of $0.958$ between subjective results and published years. SRCNN \cite{7115171} is the first CNN for super-resolution with three convolutions, and can be viewed as an end-to-end trainable sparse-coding based method \cite{yang2010image}. EDSR \cite{EDSR} adds more convolution layers with residual connections to stabilize training. DBPN \cite{DBPN} replaces single-stage upsampling with iterative up/downsampling. TSRN \cite{Texturenet} optimizes for the texture-aware LPIPS \cite{zhang2018unreasonable} metric, and underperforms RankSRGAN \cite{RankSRGAN}, ESRGAN \cite{ESRGAN}, and ESRGANplus \cite{ESRGAN+} based on GANs with stronger texture synthesis capabilities. The lastest USRGAN\cite{zhang2020deep} inherits the flexibility of model-based methods, while maintaining the end-to-end training capability of learning-based
methods.  

\begin{figure*}[t]
\begin{center}
\subfloat[USRGAN/ESRGAN]{\includegraphics[height=0.165\linewidth]{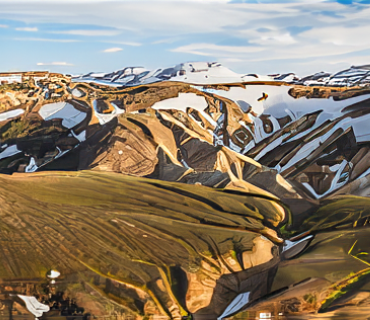}}\hskip.3em
\subfloat[USRGAN/ESRGANplus]{\includegraphics[height=0.165\linewidth]{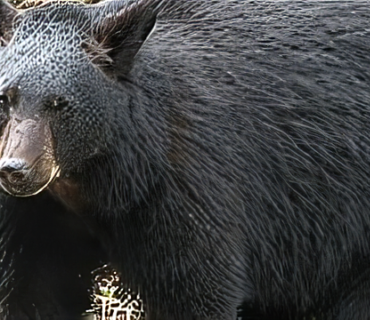}}\hskip.3em
\subfloat[RankSRGAN/ESRGAN]{\includegraphics[height=0.165\linewidth]{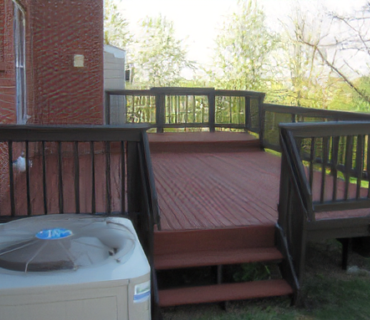}}\hskip.3em
\subfloat[TSRN/EDSR]{\includegraphics[height=0.165\linewidth]{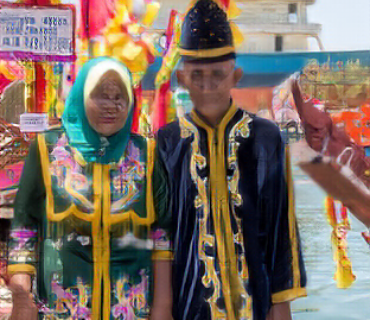}}\hskip.3em
\subfloat[TSRN/RankSRGAN]{\includegraphics[height=0.165\linewidth]{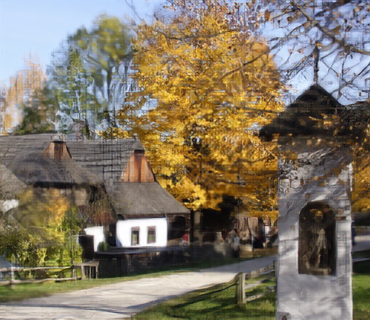}}\hskip.3em
\\
\vspace{-0.2cm}
\subfloat[DBPN/ESDR]{\includegraphics[height=0.165\linewidth]{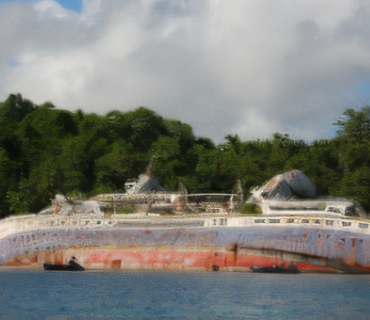}}\hskip.3em
\subfloat[DBPN/USRGAN]{\includegraphics[height=0.165\linewidth]{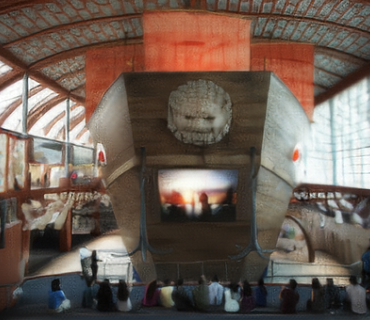}}\hskip.3em
\subfloat[RankSRGAN/ESRGAN]{\includegraphics[height=0.165\linewidth]{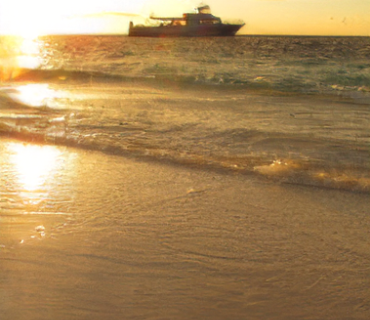}}\hskip.3em
\subfloat[EDSR/DBPN]{\includegraphics[height=0.165\linewidth]{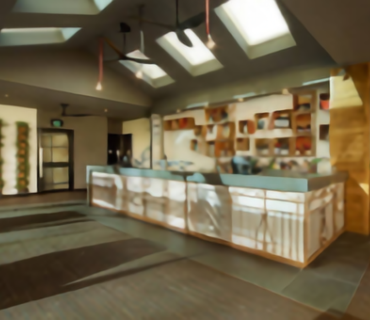}}\hskip.3em
\subfloat[SRCNN/RankSRGAN]{\includegraphics[height=0.165\linewidth]{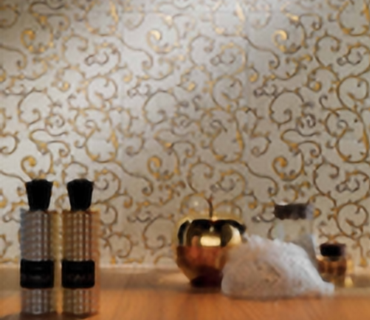}}\hskip.3em
% \fbox{\rule{0pt}{2in} \rule{.9\linewidth}{0pt}}
\end{center}
\vspace{-.5cm}
  \caption{Representative distortions created by super-resolution methods in our experiment. $f_i/f_j$ below each image means that $f_i$ is used to produce the image, while $f_j$ is the paired method for selecting the corresponding low-resolution image.}
      \vspace{-0.4cm}
\label{fig:SR}
\end{figure*}

\begin{figure*}[t]
\begin{center}
\subfloat[Fu16/BIMEF]{\includegraphics[height=0.165\linewidth]{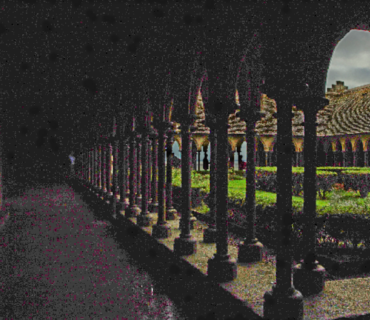}}\hskip.3em
\subfloat[Wang19/JED]{\includegraphics[height=0.165\linewidth]{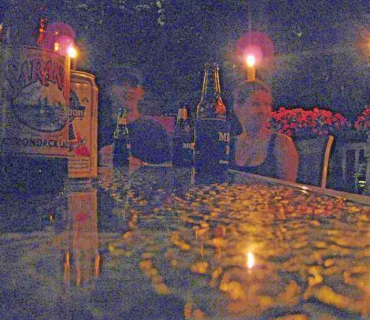}}\hskip.3em
\subfloat[Zhang19/Fu16]{\includegraphics[height=0.165\linewidth]{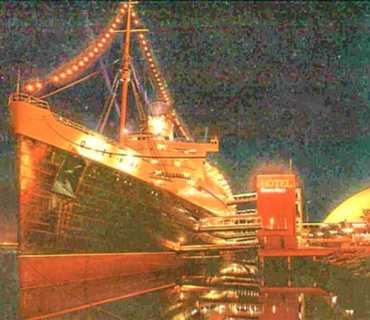}}\hskip.3em
\subfloat[Zhang19/JED]{\includegraphics[height=0.165\linewidth]{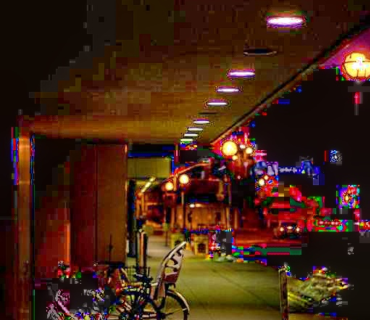}}\hskip.3em
\subfloat[Retinex-Net/JED]{\includegraphics[height=0.165\linewidth]{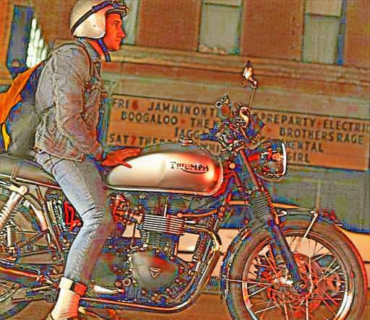}}\hskip.3em
\\
\vspace{-0.2cm}
\subfloat[Retinex-Net/Fu16]{\includegraphics[height=0.165\linewidth]{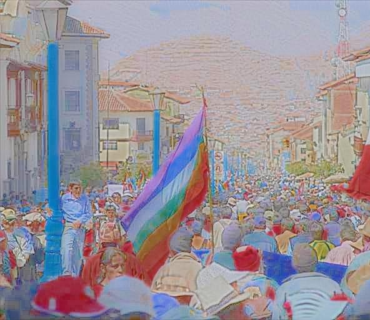}}\hskip.3em
\subfloat[EnlightenGAN/Fu16]{\includegraphics[height=0.165\linewidth]{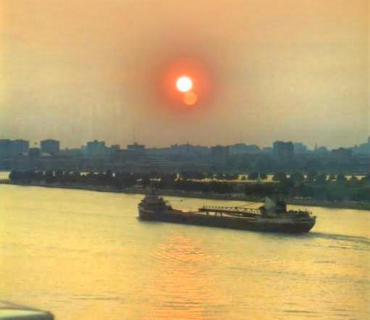}}\hskip.3em
\subfloat[JED/Zero-DCE]{\includegraphics[height=0.165\linewidth]{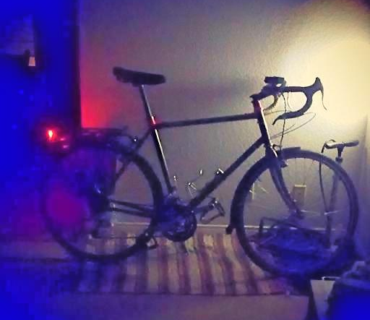}}\hskip.3em
\subfloat[BIMEF/Fu16]{\includegraphics[height=0.165\linewidth]{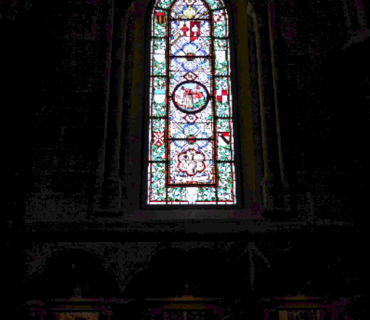}}\hskip.3em
\subfloat[Zero-DCE/Zhang19]{\includegraphics[height=0.165\linewidth]{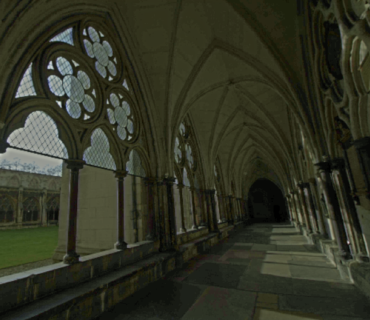}}\hskip.3em
% \fbox{\rule{0pt}{2in} \rule{.9\linewidth}{0pt}}
\end{center}
\vspace{-.5cm}
   \caption{Representative distortions created by low-light enhancers in our experiment. $f_i/f_j$ below each image means that $f_i$ is used to produce the image, while $f_j$ is the paired method for selecting the corresponding low-light image.}
    \vspace{-0.4cm}
\label{fig:enhance}
\end{figure*}

For \textit{low-light enhancement}, the main observation is that CNN-based methods have not come to dominate this field due to the lack of ground-truth normal-light images for paired supervision. For example, Retinex-Net \cite{Chen2018Retinex}, ranked in the last place, only sees $485$ realistic pairs during training, which may be insufficient to cover the real-world scene complexities. Zero-DCE \cite{Zero-DCE} optimizes a CNN for a combination of image naturalness measures, including spatial consistency, exposedness, color constancy, and illumination smoothness, without reference to normal-light images. However, the combined loss has not be calibrated against human judgments, leading to unpredictable real-world generalization. An exception is  EnlightenGAN \cite{Jiang2019enlighten}, which leverages an unsupervised GAN  to regularize the unpaired training, leading to the best performance in our subjective study. Second, it is difficult to enhance the details of low-light images without amplifying the background noise. JED \cite{Ren2018JED} performs joint optimization of low-light enhancement and noise suppression, leading to noticeable visual quality improvements. The  multi-exposure fusion framework adopted by BIMEF \cite{ying2017bio} and Zhang19 \cite{Zhang2019dual} achieves a similar effect of noise reduction with comparable performance. Third, relying on classic image processing techniques such as multi-scale decomposition and adaptive histogram equalization, Fu16 \cite{Fu2016fussion} and Wang19 \cite{Wang2019ad} tend to overshoot local details at the sacrifice of global brightness and contrast. Last, similar as dehazing, steady progress over the years is not reflected in our debiased subjective experiment, with an SRCC value of $0.071$ between the subjective results and the published years of the competing methods.

\vspace{-0.5em}
\paragraph{Qualitative Results} We show some visual examples for each of the three tasks, summarizing and diagnosing the identified distortion patterns.

% \paragraph{Selected images} We present the comparison of the selected images with the representation images of the benchmarks for the three applications in Figure\ref{fig:hazy},\ref{fig:enhancement},\ref{fig:sr}. In Figure\ref{fig:hazy}, compared to the hazy images in the benchmark, the picked images usually have multiple targets including both human objects, buildings, or vehicles with heavier fog. Also, the fog in these images is uneven and has different colors. Ambiguity introduced by motion, sandstorm, colorful artificial light, the details on the cloths and the words are also selected. From Figure\ref{fig:enhancement}, we can observe that, compared to the benchmark, the selected low-light images usually contain strong light source with halo or strong color contrast. Generally, the performance of the algorithms around the halo may be unsatisfying. Also, regarding the sky, the picked images are cloudier and monochrome. Additionally, the boundary of the objects is ambiguous. For super-resolution (Figure\ref{fig:sr}) the picked images contain more objects and straight lines.
For \textit{dehazing} in Figure \ref{fig:dehazing}, the perceived distortions can be approximately classified into five types: JPGE blocking (see  (a), (b), (c), and (d)),  color cast (see (b), (d), and (e)), loss of high-frequency information (see (f), and (g)), low-brightness (see (h) and (i)), and haze (see (i) and (j)). Knowledge-driven methods such as Dhara20 \cite{Dhara2020} and Cho18 \cite{Cho18} typically remove haze aggressively, and enhance the underlying JPEG artifacts of hazy images from the Internet accompanied by the color cast problem. Data-driven methods such as AOD-net \cite{li2017aod} and FFA-Net \cite{FFA-net} have learned a more conservative dehazing strategy. CAP \cite{zhu2015dehaze} tends to increase the global contrast of the image, leaving the dark regions darker and the hazy regions hazier. Despite the best performance, Shao20 \cite{Shao2020} is likely to smooth high-frequency details, which is successfully spotted by our debiased subjective method in (g).

For \textit{super-resolution} in Figure \ref{fig:SR}, the perceived distortions typically fall into four categories: blurring (see (f) to (j)), fake textures (see (b), (c), (d),  and (e)), incorrect semantics (see (d)), and over-enhancement of local contrast (see (a) and (b)). CNNs not optimized for texture-aware losses often suffer from blurring artifacts. CNNs reinforced by GANs are capable of synthesizing random textures, but remain weak at super-resolving structured (especially periodic) textures.
All methods fail when it comes to images with rich semantics such as faces, validating face hallucination \cite{liu2007face} as a separate super-resolution problem of its unique challenge and independent interest. With more specialized modules proposed, the field of single image super-resolution begins to enter the era of local contrast over-enhancement, as pointed out by our subjective method. 

For \textit{low-light enhancement} in Figure \ref{fig:enhance}, the perceived distortions roughly belong to five classes: noise (see (a), (b), and (c)), JPEG blocking (see (d)), abnormal brightness (see (e) and (f)), color cast (see (g) and (h)), and poor exposure (see (i) and (j)). Similar as dehazing, knowledge-driven methods (\eg, Fu16 \cite{Fu2016fussion}, Wang19 \cite{Wang2019ad}, and Zhang19 \cite{Zhang2019dual}) encourage over-enhancing details, which significantly amplifies background noise and possible JPEG blocking. Unlike super-resolution, data-driven methods (\eg, Retinex-Net \cite{Chen2018Retinex} and Zero-DCE \cite{Zero-DCE}) are far more brittle than knowledge-driven ones, which sometimes have unexpected behaviors, producing results with unnatural appearances. The best performer EnlightenGAN \cite{Jiang2019enlighten} exhibits the least amount of artifacts, but still appears to have halos around light sources in the scene, which is identified by our method.

\begin{table}[t]
  \centering
  \caption{The global ranking results of single image super-resolution under different distance measures $D_1$}
    \begin{tabular}{l|ccc}
    \toprule
      \multirow{2}{*}{Method} & 
      \multicolumn{3}{c}{Global Ranking}\\
      \cline{2-4}
    & DISTS & $\Delta$ MSE & $\Delta$ SSIM\\
     \hline 
USRGAN \cite{zhang2020deep} & 1  & 0 &  0 \\
ESRGANplus \cite{ESRGAN+} & 2  & 0  & 0 \\
ESRGAN \cite{ESRGAN} & 3  & -1 &  0  \\
RankSRGAN \cite{RankSRGAN} & 4  & +1  & 0\\
TSRN \cite{Texturenet}& 5  & 0  & -1\\
DBPN \cite{DBPN}& 6 & 0 & +1 \\
EDSR \cite{EDSR}& 7  & 0  & 0\\
SRCNN \cite{7115171}& 8  & 0 & 0\\
    \bottomrule
    \end{tabular}
  \label{tab:rank}
\end{table}

\vspace{-1em}
\paragraph{Ablation Study} We first analyze the sensitivity of our subjective results to different distance measures $D_1$ in Eq. \eqref{eq:mad2}. We use another two widely adopted metrics in signal and image processing - MSE and SSIM \cite{wang2004image}. We opt for single image super-resolution, and follow the procedure in Section \ref{subsec:setup} to sample two subsets, each of which contains $336$ pairs of images. We gather human data from $21$ subjects. Table \ref{tab:rank} shows the results, where we find that the global ranking is consistent across the three metrics. This may be because MAD chooses images to optimally discriminate between two models with large perceptual distances, which can be well approximated by all the three measures.

\begin{figure}[t]
\centering
\includegraphics[width=1\linewidth]{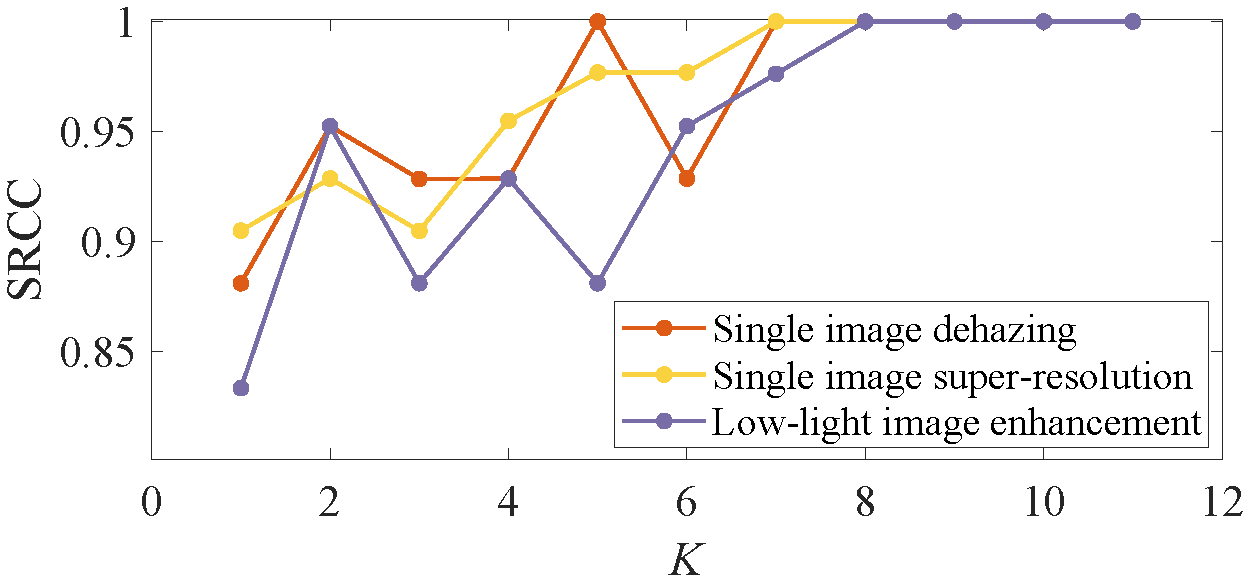}
\vspace{-0.5cm}
   \caption{The SRCC values between the top-$12$ and other top-$K$ rankings, where $K \in \{1,2,\ldots,11\}$.}
   \vspace{-0.3cm}
\label{fig:steady}
\end{figure}
Next, we analyze the sensitivity of the obtained results to $K$, \ie, the number of selected images for subjective testing. We calculate the SRCC values between the top-$12$ ranking (as reference) and other top-$K$ rankings, where $K = \{1, 2, \ldots, 11\}$. As shown in Figure \ref{fig:steady}, the ranking results are fairly stable ($\mathrm{SRCC}> 0.97$) when $K \geq 7$ for all three applications. This provides a strong indication of the sample efficiency of the proposed subjective method.

\section{Conclusion}
We have presented a debiased subjective assessment method for comparing real-world image enhancement algorithms based on the MAD competition methodology. Our method  effectively reduces the sampling, algorithmic, and subjective biases rooted in conventional subjective testing. We have demonstrated the effectiveness of the proposed method on three real-world image enhancement applications. Our method points out the caveats in the reported advances for single image dehazing and low-light image enhancement, and verifies the reliable progress in single image super-resolution with a relatively simpler degradation model. 

The application scope of the proposed debiased subjective assessment method is far beyond image enhancement. It can be broadly applied to many other subfields of computational photography, including image editing, image-to-image translation, high-dynamic-range imaging, light field imaging and more, where debiased and efficient subjective testing is largely lacking. Moreover, we may change the perceptual distances in Eq. \eqref{eq:mad2} to more general utility functions, towards benchmarking computational photography techniques for machine vision \cite{wang2020going}.
 
\section*{Acknowledgments}
The authors would like to thank all subjects who
participated in our subjective study during this period of the coronavirus pandemic. This work was supported in part by the National Natural Science Foundation of China (62071407), and the CityU SRG-Fd and APRC Grants (7005560 and 9610487).
 
% GOOD TRY TO SPOT MAD WEAKNESSES
%  First, MAD aims at relatively comparing models, and cannot give an absolute performance for only one model. Second, MAD works based on the assumption that models in the competition are reasonably good. Third, the distance in Eq. \eqref{eq:mad2} may not fully reflect human's perception of the images. Based on the above issues, the performance of MAD can be improved in the future. 
{\small
\bibliographystyle{abbrv}
\bibliography{egbib}
}

\end{document}